# Two-dimensional superconductivity and anomalous vortex dissipation in newly-discovered transition metal dichalcogenide-based superlattices

Mengzhu Shi[1‡], Kaibao Fan[1‡], Houpu Li[2], Senyang Pan[2,3], Jiaqiang Cai[2,3], Nan Zhang[2], Hongyu Li[1], Tao Wu[1,4], Jinglei Zhang[3], Chuanying Xi[3], Ziji Xiang[1,4*], Xianhui Chen[1,2,4*]

[1]Hefei National Research Center for Physical Sciences at the Microscale, University of Science and Technology of China, Hefei, Anhui 230026, China.

[2]Department of Physics and CAS Key Laboratory of Strongly-coupled Quantum Matter Physics,
University of Science and Technology of China, Hefei 230026, Anhui, China.

[3]High Magnetic Field Laboratory, Chinese Academy of Sciences, Hefei, Anhui 230031, China.

[4]Hefei National Laboratory, University of Science and Technology of China, Hefei 230088, China.

**ABSTRACT:** Properties of layered superconductors can vary drastically when thinned down from bulk to monolayer, owing to the reduced dimensionality and weakened interlayer coupling. In transition metal dichalcogenides (TMDs), the inherent symmetry breaking effect in atomically thin crystals prompts novel states of matter, such as Ising superconductivity with an extraordinary in-plane upper critical field. Here, we demonstrate that two-dimensional (2D) superconductivity resembling those in atomic layers but with more fascinating behaviours can be realized in the bulk crystals of two new TMD-based superconductors $Ba_{0.75}ClTaS_2$ and $Ba_{0.75}ClTaSe_2$. They comprise an alternating stack of H-type TMD layers and Ba-Cl layers. In both materials, intrinsic 2D superconductivity develops below a Berezinskii-Kosterlitz-Thouless transition. The upper critical field along $ab$ plane ($H_{c2}^{\|ab}$) exceeds the Pauli limit ($\mu_0 H_p$); in particular, $Ba_{0.75}ClTaSe_2$ exhibits an extremely high $\mu_0 H_{c2}^{\|ab} \approx 14\ \mu_0 H_p$ and a colossal superconducting anisotropy ($H_{c2}^{\|ab}/H_{c2}^{\perp ab}$) ~150. Moreover, the temperature-field phase diagram of $Ba_{0.75}ClTaSe_2$ under an in-plane magnetic field contains a large phase regime of vortex dissipation, which can be ascribed to the Josephson vortex motion, signifying an unprecedentedly strong fluctuation effect in TMD-based superconductors. Our results provide a new path towards the establishment of 2D superconductivity and novel exotic quantum phases in bulk crystals of TMD-based superconductors.

## INTRODUCTION

Superconductivity in 2D systems has been intensively investigated for more than 80 years due to its scientific importance as well as great promise in applicational practices[1-8]. With the dimensionality reduced from three-dimensional (3D) to 2D, numerous novel physical phenomena and phases of matter emerge, such as the Berezinskii-Kosterlitz-Thouless (BKT) transition[9,10], the superconductor-insulator quantum phase transition[8], the quantum metallic/anomalous metallic states[11,12], the quantum Griffiths phase[13], etc. Whereas the corresponding theoretical and experimental investigations have been extremely fruitful in superconducting thin films and exfoliated flakes[8], it has also been suggested that enriched 2D characteristics can be introduced to layered superconductors through sufficient weakening of interlayer coupling[4,5,14]. For instance, in layered superconductors with Josephson interaction between the superconducting planes, the orbital pair breaking effect that often prevails in the 3D superconductors is effectively suppressed for a magnetic field applied parallel to the layers[14]; subsequently, the in-plane upper critical field is predominantly determined by the paramagnetic pair breaking[15,16], giving rise to exotic phenomena including the formation of spatially inhomogeneous Fulde-Ferrell-Larkin-Ovchinnikov (FFLO) state in the clean limit[5,17].

Superconducting TMDs have long served as one of the most prominent categories of low-dimensional superconductors, owing to their inherent layered structure and large tunability upon variance of multiple parameters (layer number, strain, intercalation, electric field, twist angle, etc.)[18,19]. In recent years, atomically-thin highly-crystalline TMD superconductors have been prepared via molecular beam epitaxy and mechanical exfoliation methods, which considerably benefits the exploration of unusual superconducting properties in the 2D limit[8,18]. Distinct superconducting behaviours of TMD atomic layers and bulk crystals have been elaborately revealed; for example, strongly enhanced $H_{c2}^{\|ab}$ far beyond the Pauli paramagnetic limit $\mu_0 H_p$[15,16] has been reported in the monolayer or few-layer flakes of TMDs[20-24]. The Pauli-limit violation is associated with the Ising pairing, which originates from the breaking of inversion and in-plane mirror symmetries in monolayer TMDs[18,20-24]; such symmetry breaking effects result in a large perpendicular spin-orbit field that locks the electron spins to valley-dependent out-of-plane directions, the singlet pairing between electrons with opposite spin polarizations is thus robust against in-plane magnetic fields ($H$). Although such spin-valley locking is supposed to vanish with restored inversion symmetry in the bulk crystals[22], more recent studies figured out that once the TMD layers are effectively decoupled by intercalation of spacer layers, high $H_{c2}^{\|ab}$ due to the pairing between spin-polarized electrons stemming from Ising- or Rashba-type spin-orbit coupling (SOC) persists in the bulk[25-27]. Such realization of monolayer characteristics in macroscopic crystals promises a better chance for the research of 2D superconductivity, because bulk samples have more advantages compared with the atomic layers, e.g., low air sensitivity, free of sophisticated fabrication processes and a much larger applicability range of measurement techniques.

In this work, we report the discovery of TMD-based superconductors $Ba_{0.75}ClTaS_2$ and $Ba_{0.75}ClTaSe_2$, whose crystal structure hosts alternating stacked H-type TMD and Ba-Cl layers. Superconducting transition occurs at 2.75 K and 1.75 K in $Ba_{0.75}ClTaS_2$ and $Ba_{0.75}ClTaSe_2$, respectively. A comprehensive study of their electrical transport and magnetic properties unveils pronounced 2D characteristics for the superconductivity in single crystals of these superlattices. Most strikingly, $Ba_{0.75}ClTaSe_2$ exhibits an extreme $\mu_0 H_{c2}^{\|ab}$ of ~ 50 T, one order of magnitude higher than $\mu_0 H_p$, hinting at the emergence of Ising pairing resulted from interlayer decoupling; concomitantly, the flux-flow resistance develops an unexpectedly complex evolution under magnetic field, implying peculiar processes of Josephson vortex (JV) depinning and motion stemming from strong superconducting fluctuations. Our observations explicitly demonstrate that TMD-based superlattices are not only good candidates for realizing 2D superconductivity in bulk crystals, but also a fertile playground for exploring the enriched phenomena that have been extensively linked to unconventional superconducting pairing.

**EXPERIMENTAL SECTION**

The synthesis of the single crystal $Ba_{0.75}ClTaS_2$ and $Ba_{0.75}ClTaSe_2$ is achieved through a flux method. Pieces of Ba (99.5%), Ta powder (99.5%), S or Se powder (99.999%), and anhydrous $BaCl_2$ powder (99.9 %) were mixed with the molar ratio of 0.6:1:2:25 in a glove box filled with Argon gas. The atmosphere in the glove box guarantees oxygen and water levels below 1 ppm; to avoid unintentional water absorption, the $BaCl_2$ powder was first heated at 700 °C for 20 h before transferring into the glove box. The mixture was loaded into an alumina crucible, which was placed in a quartz ampoule. The ampoule was subsequently placed in a large quartz tube to form a double-walled assembly. Such assembly was then evacuated, sealed, placed in a furnace and heated to 1150 °C in 24 h. After keeping at this temperature for 24 h, the furnace was slowly cooled down to 800 °C in 7 days, followed by a power-off cooling. Single crystals with typical dimensions of $0.2 \times 0.4 \times 0.02$ mm$^3$ were obtained after washing the product using dimethyl formamide (99.9%) for 2 days in a glove box. The specific characterization details including structure characterization, transport measurement and magnetization measurements are shown in the supporting information.

**RESULTS AND DISCUSSION**

**Structure of the superlattices.** Single crystals of $Ba_{0.75}ClTaS_2$ and $Ba_{0.75}ClTaSe_2$ were grown by the flux method (see supporting information for details). To determine the crystal structure, we conducted powder X-ray diffraction (pXRD), four-circle single crystal X-ray diffraction (SC-XRD), high-angle annular dark field (HAADF) imaging, selected area electron diffraction (SAED) and energy dispersive X-ray spectrum (EDX) measurements; the results are summarized in Figure 1 (see Figure S1, Tables S1 and S2). Both materials adopt a previously unreported structure with monoclinic symmetry (space group C2/c, No.15). As shown in the Figure 1a, this structure consists of alternating stack of H-$TaS_2$/H-$TaSe_2$ layers and Ba-Cl spacer layers along the crystallographic $c$ axis. In the H-type TMD layers, each Ta atom forms chemical bonds with six S or Se atoms, resulting in a triangular-prism coordination polyhedron. The Ba-Cl layers incorporate disordered vacancies; we note that the Wyckoff sites of Ba and Cl are close to each other and the sum of the atomic occupancy of these two sites appears to be less than 1 (Table S2). The lattice parameters of the superlattices (see Table S1) approximately satisfy the following relationships: $a = a_0$, $b = \sqrt{3}a_0$, $0.5 \times c \times \sin(\beta) = c_0 + c_1$, $a \times b = a_1 \times a_1$, where $a_0$ and $c_0$ are the lattice parameters of the TMD layer, $a_1$ and $c_1$ are the in-plane lattice parameter and thickness of Ba-Cl sub-layer in compound BaClF[28], respectively. If the partial occupation in Ba-Cl layers is ignored, such relationships correspond to the composition $Ba_2Cl_2Ta_2(S/Se)_4$ for the superlattice unit cell illustrated in Figure 1a; the commensurate stacking of the two sub-layers is corroborated by the SAED image (Figure S1c). The actual chemical formula determined by SC-XRD refinement (Table S1) and the EDX results (Figure 1f) turns out to be $Ba_{0.75}ClTa(S/Se)_2$. High quality of single crystals is confirmed by the sharp spots reconstructed from the SC-XRD (Figure 1b and c) and the sharp *(00l)* diffraction peaks (Figure 1e). Cross-section HAADF image (Figure 1d) for $Ba_{0.75}ClTaS_2$ shows a clear alternating stack of H-$TaS_2$ and Ba-Cl layers with no visible stacking faults. Owing to the presence of Ba-Cl layers, the spacing between adjacent TMD layers is 12.03 Å (12.32 Å) for $Ba_{0.75}ClTaS_2$ ($Ba_{0.75}ClTaSe_2$), remarkably expanded comparing with 2H-$TaS_2$ and 2H-$TaSe_2$ (5.86 Å and 6.35 Å, respectively). Similar intergrowth of metal halide sub-layer has also been realized in transition metal oxides $Sr_2Cl_2CuO_2$[29], which is the high-$T_c$ cuprate superconductor with single $CuO_2$ plane in one unit cell, and $Sr_2Cl_2NiO_2$[30].

**Resistivity and superconducting transition.** Expansion of the distance between conducting TMD layers naturally renders the electron system more two dimensional, resulting in strong anisotropy in electrical transport properties. This is demonstrated by the temperature ($T$) dependence of in-plane ($\rho_{ab}$) and out-of-plane ($\rho_\perp$) resistivity data displayed in Figure 2a and b. For both compounds, $\rho_{ab}$ exhibits metallic behaviour which decreases upon cooling and saturates below ~ 20 K. By contrast, $\rho_\perp$ measured in $Ba_{0.75}ClTaS_2$ is insulating-like with a negative $d\rho_\perp/dT$ over the entire temperature range (Figure 2a); the resistivity anisotropy $\rho_\perp / \rho_{ab}$ reaches approximately 2400 at low *T* (Figure S2a). For $Ba_{0.75}ClTaSe_2$, the resistivity anisotropy is somewhat weaker (~1060 at the lowest *T*, Figure S2b) and $\rho_\perp$ exhibits a nonmonotonic (albeit weak) *T* dependence. The nonmetallic interlayer charge transport implies that the Fermi surfaces of these materials are strongly 2D and, apart from the normal coherent (Boltzmann) relaxation, the interlayer hopping due to tunneling through resonant levels of impurities (such as Ba vacancies in the spacing layers) ought to be considered[31]. Note that in the parent compounds 2H-$TaS_2$[32] and 2H-$TaSe_2$[33], the *c*-axis resistivity remains metallic and the resistivity anisotropies are a little lower than our $TaS_2$/$TaSe_2$-based superlattices (Table S3). Hall resistivity measurements reveal dominant hole-type carriers in the superlattices (Figure S3a and b).

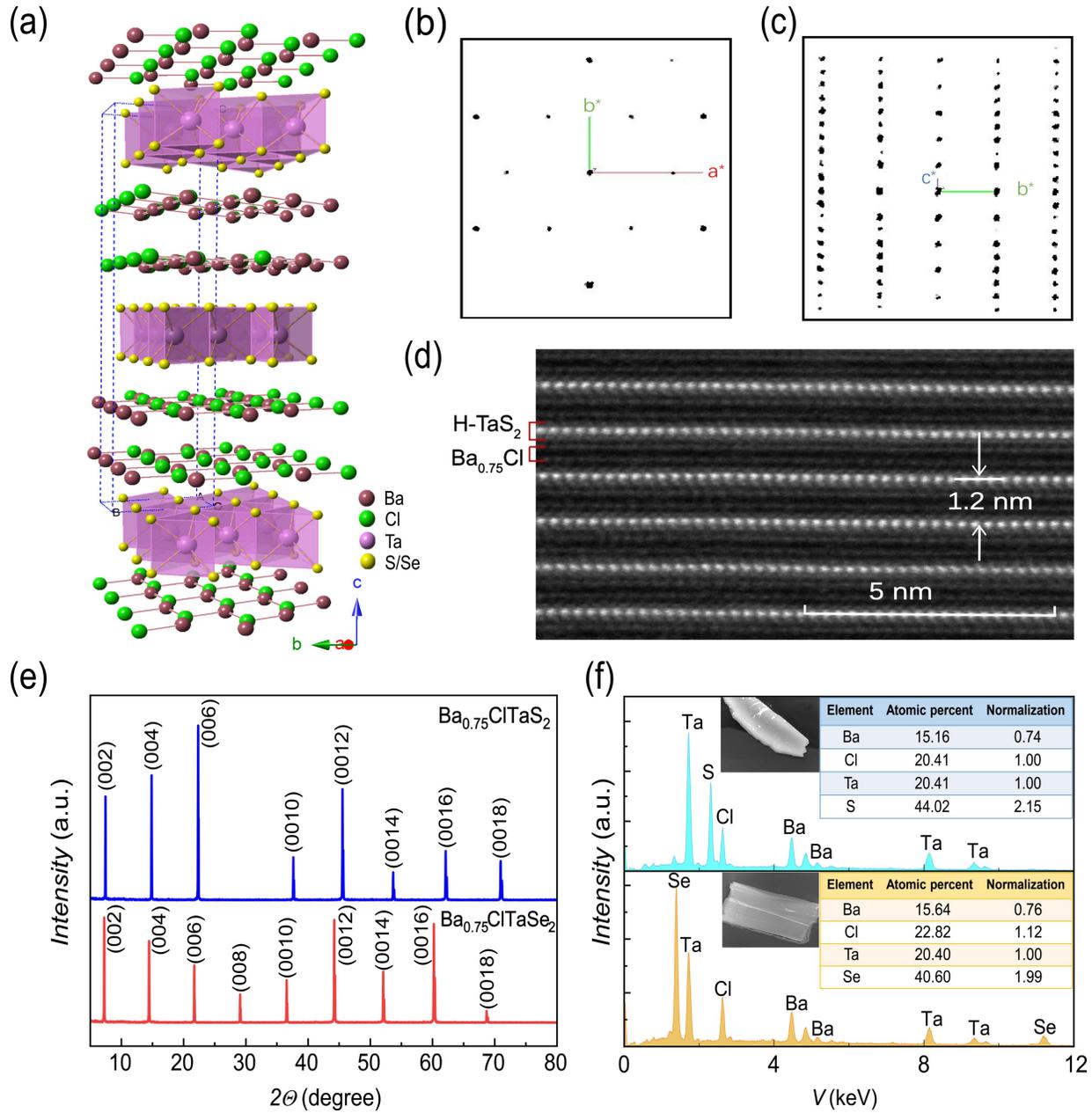

**Figure 1.** Structural characterizations of $Ba_{0.75}ClTaS_2$ and $Ba_{0.75}ClTaSe_2$. (a) Illustration of the crystal structure, which is composed of alternating stacked H-TMD and Ba-Cl layers. Blue dashed lines denote the superlattice unit cell. (b, c) Reconstructed X-ray diffraction (XRD) spots of a $Ba_{0.75}ClTaS_2$ single crystal, viewed from *c*-axis (b) and *a*-axis (c), respectively. We note the average mosaicity, which describes the stacking quality of the small domain in a crystal, is approximately 1.5°, 1° and 1° along three axes, respectively. (d) A cross-section HAADF image measured in a $Ba_{0.75}ClTaS_2$ single crystal. (e) Single crystal (SC-)XRD patterns for $Ba_{0.75}ClTaS_2$ and $Ba_{0.75}ClTaSe_2$; (00*l*) series of diffraction peaks are clearly observed. (f) The EDX spectrum for $Ba_{0.75}ClTaS_2$ and $Ba_{0.75}ClTaSe_2$. The inset shows the SEM image and chemical composition of the corresponding material. The element ratio Ba : Cl : Ta : S(Se) is close to 0.75:1:1:2, with Ba : Cl : Ta : S = $(0.74 \pm 0.01) : (1.00 \pm 0.02) : 1 : (2.15 \pm 0.04)$ and Ba : Cl : Ta : Se = $(0.76 \pm 0.01) : (1.12 \pm 0.03) : 1 : (1.99 \pm 0.05)$.

Determined from the Hall data taken at 5 K (2.5 K), the Hall carrier density $n_H$ and Hall mobility $\mu_H$ can be estimated: $n_H$ = 0.6(0.88) × $10^{21}$ $cm^{-3}$, $\mu_H$ = 150(60) $cm^2V^{-1}s^{-1}$ for $Ba_{0.75}ClTaS_2$ ($Ba_{0.75}ClTaSe_2$) (see Figure S3e and f for additional data). A weak nonlinearity of $\rho_{xy}(H)$ is observed in $Ba_{0.75}ClTaSe_2$ at 2.5 K, implying two-band transport at low $T$ (supporting information Figure S4). We adopt a single band discussion about the transport behavior at normal state in the paper for simplicity and the conclusions are not affected in the two band model. Below $T \simeq 15$ K, the low-field ($\mu_0 H < 0.2$ T) magnetoresistance (MR) is featured by a small dip centered at $H = 0$ for the both superlattices (Figure S3c and d); such phenomenon is a clear manifestation of weak anti-localization (WAL) effect[34], corroborating the quasi-2D nature of charge transport in these samples. More specific WAL analysis is shown in the supporting information (Figure S5).

The most interesting aspect of the new TMD-based superlattices is the emergence of superconductivity at low $T$. The superconducting transition temperature $T_{c0}$ defined as the onset of zero resistivity is 2.75 K (1.75 K) for $Ba_{0.75}ClTaS_2$ ($Ba_{0.75}ClTaSe_2$) (insets of Figure 2a and b); $T_{c0}$ is consistent with the temperature below which diamagnetic signal due to the Meissner effect develops on the magnetic susceptibility (Figure 2c and d). The zero-field-cooled susceptibility curves indicate a shielding volume fraction close to 100% in both materials, implying bulk superconductivity. The much higher $T_c$ in the TMD-based superlattices compared with 2H-TaS$_2$ and 2H-TaSe$_2$ ($T_c$ = 1.0 K[35] and 0.2 K[36], respectively) is probably attributed to the suppression of competing electronic order, i.e., the charge density wave (CDW) order that prevails in the parent compounds. In $Ba_{0.75}ClTaS_2$ and $Ba_{0.75}ClTaSe_2$, no obvious signatures of CDW transition can be identified on the electrical resistivity (Figure 2a-b and Figure S6) as well as in the Raman spectra up to 300 K, magnetic susceptibility, specific heat and SAED data (Figure S7); considering that the CDW order causes deplet-ion of density of states at Fermi energy, we conjecture that a higher $T_c$ is probably associated with its absence. More discussion on the relationship between the $T_c$ and CDW is shown in the supporting information.

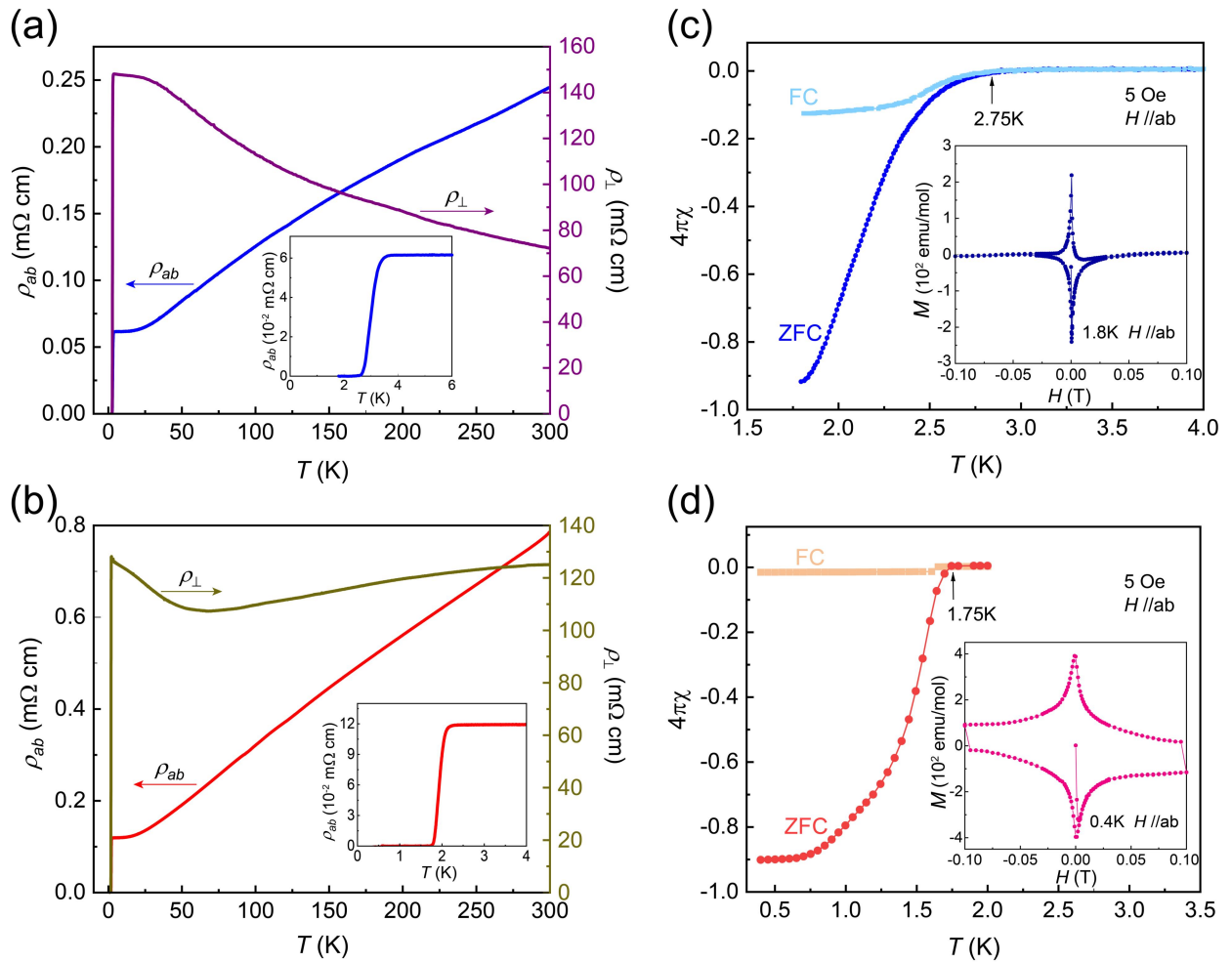

**Figure 2.** Resistivity and superconducting transition in $Ba_{0.75}ClTaS_2$ and $Ba_{0.75}ClTaSe_2$. (a, b) $T$-dependent in-plane resistivity ($\rho_{ab}$) and out-of-plane resistivity ($\rho_\perp$) measured in $Ba_{0.75}ClTaS_2$ (a) and $Ba_{0.75}ClTaSe_2$ (b), respectively. Insets show expanded views of the superconducting transition. (c, d) Magnetic susceptibility as a function of $T$ measured in an in-plane magnetic field of 5 Oe for $Ba_{0.75}ClTaS_2$ (c) and $Ba_{0.75}ClTaSe_2$ (d), respectively. Zero-field cooled (ZFC) and field-cooled (FC) curves are displayed in different colors. Insets display the magnetic hysteresis loop in the superconducting state, measured under in-plane magnetic fields at 1.8 K and 0.4 K for $Ba_{0.75}ClTaS_2$ and $Ba_{0.75}ClTaSe_2$, respectively.

**The 2D superconductivity**. Since the normal state electrical transport is highly anisotropic in both $Ba_{0.75}ClTaS_2$ and $Ba_{0.75}ClTaSe_2$, we anticipate that their superconducting states show remarkable 2D characteristics. In the 2D limit, the superconducting transition is associated with the dissociation of thermally activated vortex-antivortex pairs[37], in analogy with the BKT transition in superfluid $^4$He films[38]; below the transition temperature $T_{BKT}$, a nonlinear current-voltage ($I$-$V$) relationship $V \propto I^\alpha$ develops[39,40] with the exponent $\alpha(T)$ taking the value $\alpha = 3$ at $T = T_{BKT}$. We testified such power-law behaviour of the $I$-$V$ characteristics in the vicinity of superconducting transition in $Ba_{0.75}ClTaS_2$ and $Ba_{0.75}ClTaSe_2$. As shown in Figure 3a and b, for both compounds, the $I$-$V$ characteristics exhibit a crossover from linear (i.e., $\alpha = 1$) in the normal state to strongly nonlinear ($V \propto I^\alpha$, $\alpha \gg 1$) as $T$ decreases across the superconducting transition; the BKT transition determined by $\alpha = 3$ occurs at $T_{BKT}$ = 2.56 K (1.67 K) for $Ba_{0.75}ClTaS_2$ ($Ba_{0.75}ClTaSe_2$) (Figure 3c and d). Moreover, the resistivity immediately above $T_{BKT}$ is given by $\rho(T) = \rho_0 \exp[-b/(T-T_{BKT})^{1/2}]$ in 2D superconductors, where $\rho_0$ and $b$ are material-dependent parameters[39,40]. Accordingly, a linear fit of $[d\ln\rho/dT]^{-2/3}$ as a function of $T$ yields another criterion for the determination of $T_{BKT}$, i.e., the temperature at which the $T$-linear extrapolation crosses zero[41]. This approach provides $T_{BKT}$ = 2.54 K (1.76 K) for $Ba_{0.75}ClTaS_2$ ($Ba_{0.75}ClTaSe_2$) (Figure S9), consistent with the values given by the $I$-$V$ analysis. Thereby, 2D superconducting states are likely to be realized in bulk $Ba_{0.75}ClTaS_2$ and $Ba_{0.75}ClTaSe_2$ below a BKT transition that approximately coincides with the zero-resistance temperature $T_{c0}$. The 2D superconductivity in $Ba_{0.75}ClTaS_2$ is further supported by the angular dependence of upper critical field, $H_{c2}$, as shown in Figure 3e. At $T$ = 2.5 K, $H_{c2}$ decreases rapidly when $H$ is inclined with respect to the $ab$ plane (Figure S10a), resulting in a cusp-like profile at $\theta = 90°$ ($\theta$ is the tilt angle between $H$ and the normal vector of TMD layers); this sharp feature is a hallmark of 2D superconductivity. Fits to the 2D Tinkham formula[42] provide a good description of the $H_{c2}(\theta)$ profile, whereas the 3D Ginzburg-Landau (GL) model works less satisfyingly (curves in Figure 3e, see supporting information for details).

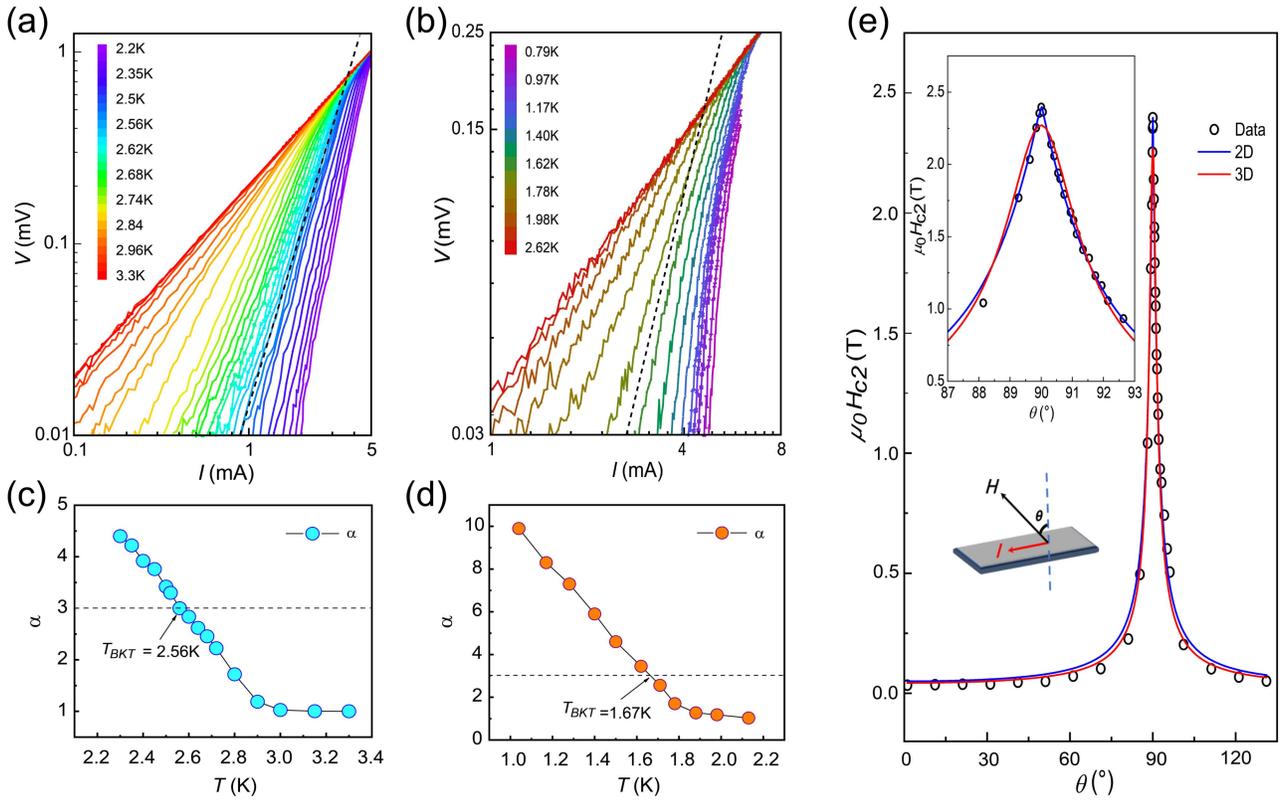

**Figure 3.** The 2D superconductivity in $Ba_{0.75}ClTaS_2$ and $Ba_{0.75}ClTaSe_2$. (a, b) Current-voltage ($I$-$V$) characteristics at various temperatures plotted in a logarithmic-logarithmic scale for $Ba_{0.75}ClTaS_2$ (a) and $Ba_{0.75}ClTaSe_2$ (b). The black dash lines indicate the power-law behaviour $V \propto I^3$. (c, d) Temperature dependence of the power-law exponent $\alpha$ deduced from the $V \propto I^\alpha$ relationship in $Ba_{0.75}ClTaS_2$ (c) and $Ba_{0.75}ClTaSe_2$ (d). The BKT transition temperature $T_{BKT}$ defined as the temperature at which $\alpha = 3$ is 2.56 K and 1.67 K for $Ba_{0.75}ClTaS_2$ and $Ba_{0.75}ClTaSe_2$, respectively. (e) Angular dependence of the upper critical field $H_{c2}$ in $Ba_{0.75}ClTaS_2$, which is measured at 2.5 K (below $T_{c0}$). $H_{c2}$ in this plot is determined as the magnetic field at which the resistivity reaches 50% of its normal state value (see Figure S10a for raw data); $\theta$ is the angle between the magnetic field vector and the normal direction of the TMD layer ($ab$ plane). The electrical current was applied in-plane. Red and blue curves represent fittings of $H_{c2}(\theta)$ based on 3D Ginzburg-Landau model and Tinkham's 2D model, respectively (see supporting information). The inset shows the enlarged view in the close vicinity (±3°) of $\theta$ = 90°.

**Upper critical fields.** Behaviours revealed by BKT and $H_{c2}(\theta)$ analysis mimic those observed in superconducting heterointerfaces[41,43], which are exemplary 2D superconductors. Such coincidence points towards a scenario that superconductivity is largely confined within the TMD layer, whilst the interlayer coupling is weak. The 2D nature of superconducting states also manifests itself in a large superconducting anisotropy $\gamma = H_{c2}^{\|ab}/H_{c2}^{\perp ab}$, where $H_{c2}^{\perp ab}$ and $H_{c2}^{\|ab}$ are upper critical fields perpendicular and parallel to the $ab$-plane, respectively. Results of upper critical field study in Ba$_{0.75}$ClTaS$_2$ are presented in Figure 4a-c and Figure S10b. The profiles of $H_{c2}(T)$ along both directions exhibit an upward curvature, which can be attributed to the multiband effect: a two-band model considering both orbital and paramagnetic pair breaking (see supporting information for details) yields reasonable fits to all the data (Figure 4c, Figure S10b). We mention that the superconducting transition in resistivity is notably broadened under $H$ (Figure 4a and b), suggesting that the apparent boundary between superconducting and normal states is smeared by strong fluctuation and/or flux-flow resistance. In this sense, we define $H_{c2}(T)$ to be the field where the resistivity recovers 95% of its normal state value. Using this criterion, zero-temperature upper critical fields of Ba$_{0.75}$ClTaS$_2$ estimated from the two-band model are $\mu_0 H_{c2}^{\perp ab}(0) \cong 0.93$ T and $\mu_0 H_{c2}^{\|ab}(0) \cong 14.93$ T, corresponding to $\gamma = 16.1$. A larger $\gamma \approx 22.7$ can be obtained if $H_{c2}$ is determined using the criterion of 50% normal state resistivity (Figure S10b). These values of $\gamma$ are comparable with other TMD-based superconductors wherein 2D superconductivity has been reported, such as Ba$_6$Nb$_{11}$(S/Se)$_{28}$[26,44] and 4Hb-TaS$_2$[45] (Table S3). Moreover, the estimated $H_{c2}^{\|ab}(0)$ appears to be considerably higher than the weak-coupling Pauli paramagnetic limit $\mu_0 H_p$ ($\approx 1.84 \times T_c$[15,16] = 6.2 T, green dotted line in Figure 4c), highlighting the influence of spin-orbit interactions.

More intriguingly, the superconducting state in Ba$_{0.75}$ClTaSe$_2$ exhibits a plethora of peculiar characteristics, suggestive of further enhanced 2D character comparing with Ba$_{0.75}$ClTaS$_2$. According to the electrical transport measurements (Figure 4d-f and Figure 5b), zero-temperature upper critical fields of Ba$_{0.75}$ClTaSe$_2$ are estimated to be $\mu_0 H_{c2}^{\perp ab} \cong 0.35$ T and $\mu_0 H_{c2}^{\parallel ab} \cong 52$ T (Figure 4f). (Here $\mu_0 H_{c2}$ is determined using the 95% threshold. More discussion about the fitting with different criterion and different model is shown in the supporting information Figure S11, S12 and Table S5) This yields an extremely large superconducting anisotropy $\gamma \approx 150$, far exceeding that of any known TMD-based superconductors (Table S3). Whereas $\mu_0 H_{c2}$ along both directions can also be fitted to our two-band model (dashed lines in Figure 4f, see supporting information for details), it should be pointed out that the abnormally high $\mu_0 H_{c2}^{\parallel ab}$ overcomes the weak-coupling Pauli paramagnetic limit $\mu_0 H_p$ ($\approx 1.84 \times T_c = 3.8$ T) by almost 14 times. Such a radical Pauli limit violation cannot be ascribed to spin-orbit scattering[14] because it requires an unphysically short scattering time (< 10 fs, see supporting information). Alternatively, we assign the high $H_{c2}^{\parallel ab}$ to the putative local symmetry breaking in Ba$_{0.75}$ClTaSe$_2$: despite that the global inversion symmetry is preserved in the crystal lattice (Figure 1a), the remarkably weakened interlayer coupling and the consequently amplified two dimensionality render the electronic structure similar to that of monolayer H-TaSe$_2$, which lacks an inversion center[26]; in such case, $H_{c2}^{\|ab}$ can still be significantly enhanced due to Ising-type SOC[25,27]. The same arguments may also be applied to Ba$_{0.75}$ClTaS$_2$, though the smaller $H_{c2}/H_p$ ratio makes the conclusion less unambiguous. More discussion about the possible contribution of Ising SOC to the enhanced upper critical field is shown in the supporting information (Figure S13).

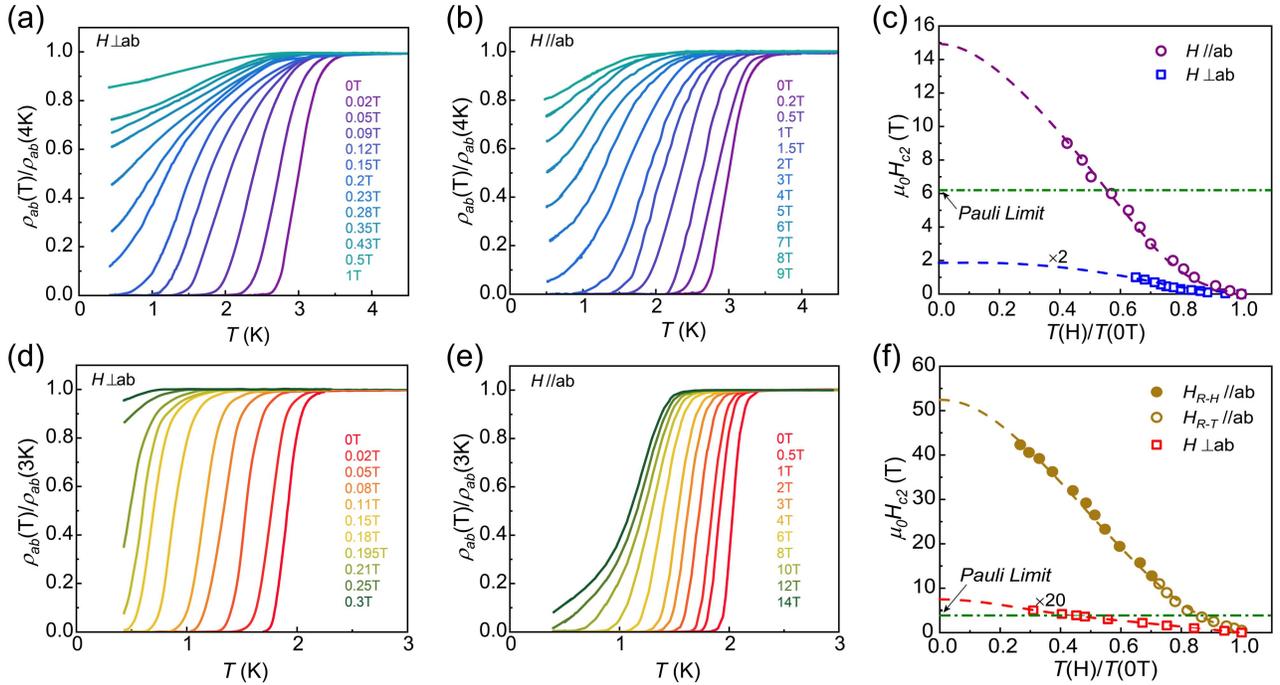

**Figure 4.** Upper critical fields in Ba$_{0.75}$ClTaS$_2$ and Ba$_{0.75}$ClTaSe$_2$. (a, b) Resistivity $\rho(T)$ measured between 0.4 K and 4.5 K in Ba$_{0.75}$ClTaS$_2$ under magnetic fields applied perpendicular (a) and parallel (b) to the ab plane. $\rho(T)$ is normalized to the value at 4 K. (c) Upper critical fields plotted against normalized temperature $T/T_c$ for Ba$_{0.75}$ClTaS$_2$. Blue and purple symbols represent $H_{c2}$ measured with $H$ perpendicular and parallel to the $ab$ plane, respectively; the former ($H_{c2}^{\perp ab}$) are multiplied by 2 for clarity. (d, e) Resistivity $\rho(T)$ measured between 0.4 K and 3 K in Ba$_{0.75}$ClTaSe$_2$ under magnetic fields applied perpendicular (d) and parallel (e) to the $ab$ plane up to 14 T. $\rho(T)$ is normalized to the value at 3 K. (f) Upper critical fields plotted against normalized temperature $T/T_c$ for Ba$_{0.75}$ClTaSe$_2$. Red and dark yellow symbols represent $H_{c2}$ measured with $H$ perpendicular and parallel to the $ab$ plane, respectively; the former ($H_{c2}^{\perp ab}$) are multiplied by 20 for clarity. Solid and hollow symbols are data taken in a hybrid magnet up to 43 T (see Figure 5b) and in a superconducting magnet up to 14 T, respectively. In (c) and (f), $H_{c2}$ is defined to be when $\rho(H)$ reaches 95% of its normal state value, $T_c$ is determined using the same threshold; dashed curves are fits to a two-band model (supporting information); green dotted lines denote the Pauli paramagnetic limit $\mu_0 H_p = 1.84 \times T_c$.

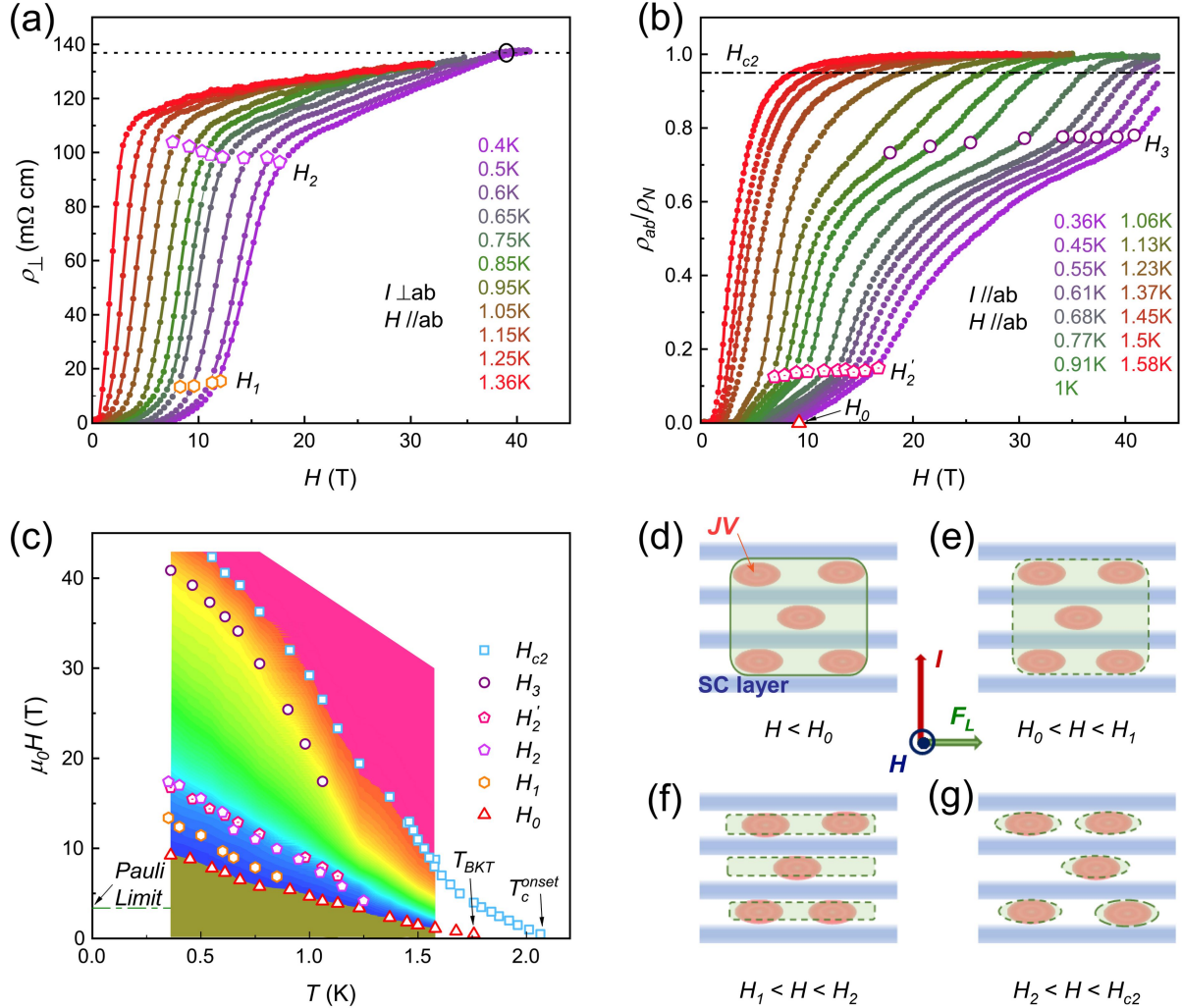

**Figure 5.** Superconducting phase diagram and Josephson vortex dynamics in Ba$_{0.75}$ClTaSe$_2$. (a, b) $H$-dependent $\rho_\perp$ (a) and $\rho_{ab}$ (b) measured in Ba$_{0.75}$ClTaSe$_2$ with $H$ applied parallel to the $ab$ plane. Resistivity curves are normalized to the normal state value in (b). Characteristic magnetic fields $H_1$, $H_2$, $H_2'$, $H_3$ are defined by the abrupt changes in the slope of resistivity (Figure S15). The black circle in a represents a possible superconducting-to-normal transition (supporting information and Figure S15a). (c) Field-temperature ($H$-$T$) phase diagram of the superconducting state in Ba$_{0.75}$ClTaSe$_2$ under an in-plane $H$. Profiles of all the $T$-dependent characteristic fields (open symbols) are overlayed onto the contour plot of normalized $\rho_{ab}(H)$ (data presented in b). The upper critical field $H_{c2}$ is defined to be when $\rho(H)$ reaches 95% of its normal state value dash-dotted line in b. $H_0$ is determined by the offset of the zero-resistance state. The Pauli paramagnetic limit $\mu_0 H_p$ is denoted by a green dotted line. (d-g) A putative scenario of the motion of JVs (red ellipsoids) driven by the Lorentz force $F_L$ (green arrow) under the application of in-plane field and out-of-plane current. The blue bars represent the superconducting (SC) layers. Green shaded areas signify whether the JVs are collectively or individually driven; the solid and dashed boundaries denote pinned and depinned JVs, respectively. (d) The pinning of JV lattice persists up to $H_0$, at which $F_L$ eventually overcomes the pinning barrier. (e) Between $H_0$ and $H_1$, JVs are depinned and driven into a collective motion. (f) At $H = H_1$ the JVs are decoupled between adjacent layers, each layer moves separately driving by $F_L$. (g) Above $H_2$, JV lattice completely melts into individual vortices.

**Superconducting phase diagram and Josephson vortex dynamics.** Based on the anisotropic GL theory[42,46], the coherence lengths of Ba$_{0.75}$ClTaSe$_2$ can be determined from the upper critical fields: $\mu_0 H_{c2}^{\perp ab} = \phi_0/(2\pi\xi_{ab}^2)$, $\mu_0 H_{c2}^{\|ab} = \phi_0/(2\pi\xi_{ab}\xi_\perp)$, here $\xi_{ab}$ and $\xi_\perp$ are the in-plane and out-of-plane coherence lengths, respectively. We estimated zero-temperature coherence lengths to be $\xi_{ab} \approx 30$ nm and $\xi_\perp \approx 0.2$ nm. Given that $\xi_\perp$ is considerably shorter than the inter-(TaSe$_2$)-layer distance $d = 1.23$ nm (Table S4; in comparison, $\xi_\perp$ and $d$ are comparable in Ba$_{0.75}$ClTaS$_2$), the GL description is indeed inappropriate for Ba$_{0.75}$ClTaSe$_2$ and this material is better regarded as a stack of 2D superconducting planes with interlayer Josephson coupling[47,48]. As the superconducting order parameter is suppressed to zero between adjacent TaSe$_2$ layers, an external magnetic field applied parallel to the plane can penetrate the interlayer space (the insulating Ba-Cl blocks in the present case), leading to the formation of JVs[4,47,49-51]. The relatively weak pinning of these Josephson flux lines results in unique vortex dissipation, which we scrutinized in Ba$_{0.75}$ClTaSe$_2$ up to 43 Tesla in a hybrid magnet at the Chinese High Magnetic Field Laboratory in Hefei. We note that the $M$-$H$ curve for Ba$_{0.75}$ClTaSe$_2$ at 0.4K (Figure 2d) indicates that the pinning is obvious in the SC layer. As revealed by high-$H$ MR measurements (results shown in Figure 5a and b), the flux-flow resistivity displays a surprisingly enriched cascade of features inside the superconducting state. The zero-resistance state terminates at a field $H_0$ much lower than $H_{c2}^{\|ab}$; we identify this field as the occurrence of depinning of the JVs (note that $H_0$ locates above the irreversibility field $H_{irr}$ determined from magnetic torque measurements, see supporting information and Figure S14 for details). Between $H_0$ and $H_{c2}^{\|ab}$, two kinks are observed in the interlayer resistivity $\rho_\perp$ at characteristic fields $H_1$ and $H_2$, respectively (Figure 5a and Figure S15a). Concurrently, the intraplane $\rho_{ab}$ shows two kinks at $H_2'$ and $H_3$, respectively (Figure 5b and Figure S15b). The $T$ dependence of all these characteristic fields is summarized in Figure 5c; the coincidence between $H_2$ and $H_2'$ suggests the same underlying mechanism for them. We note that similar flux-flow resistivity behavior is observed in other samples (Figure S16). We mention that there is no anomaly in the $H$-$T$ profiles that can signal an FFLO state[5,17] or other field-induced superconducting phases.

The complex flux-flow transport underscores the intricate nature of Lorentz-force-driven JV motion. With the application of an in-plane $H$ and a perpendicular $I$, a Lorentz force $F_L$ is exerted on the JV lattice, as shown in Figure 5d. Above the depinning line $H_0(T)$, JVs are collectively driven by $F_L$ into a horizontal motion (Figure 5e), giving rise to a voltage signal in the interlayer direction, i.e., a finite $\rho_\perp$. At $H = H_1$, $\rho_\perp$ starts a rapid increase (Figure 5a). We postulate that $H_1$ represents a characteristic field at which the JVs are decoupled between neighboring layers: above $H_1$, flux lines in each layer can be driven by $F_L$ independently (Figure 5f) with a reduced pinning barrier, therefore enhances $\rho_\perp$ [52,53]. On the other hand, $H_2$ may be associated with the onset of a true liquid state of JVs, wherein each vortex is individually driven (Figure 5g); $\rho_\perp$ reaches roughly 75% of the normal state value at this field (Figure 5a). The resistive kinks appearing in $\rho_{ab}$ (Figure 5b) are less well understood at this stage: since the motion of JVs is confined in a lateral channel between layers and thus can only be effectively driven by a perpendicular $I$[54] (Figure 5d-g), their dynamics is not directly linked to in-plane energy dissipation. One possibility is that due to a small sample misalignment, the applied $H$-field is slightly tilted from the $ab$ plane, thus the JVs penetrate the superconducting planes and 2D pancake vortices (stemming from the perpendicular component of $H$) are created within the superconducting layers; the motion of these pancake vortices[47,48,51,54,55] contributes to the complicated high-$H$ evolution of $\rho_{ab}$. The intimate interplay between ensembles of Josephson and pancake vortices ensures the manifestation of $H_2'$ feature in $\rho_{ab}$ (Figure 5b), which is likely to reflect a change of pancake vortices dissipation linked to the eventual JV melting at $H_2$.

The remarkably large flux-flow regime in the $H$-$T$ phase diagram delineated in Figure 5c is highly unusual for a TMD-based superconductor. Instead, it resembles the superconducting phase diagrams of high-$T_c$ copper oxide superconductors[48] and quasi-2D organic superconductors[56] --- both are endowed with a depinning line much below $H_{c2}$ and a broad vortex liquid state. In essence, these are hallmarks of a vortex lattice with large mean-field fluctuation amplitudes, unequivocally pointing towards strong thermal and/or quantum fluctuation effects[48,57]. The importance of quantum fluctuation in Ba$_{0.75}$ClTaSe$_2$ is verified by the ratio $Q$ between normal state resistance and quantum resistance: $Q = (e^2/\hbar)\rho_{ab}/d \approx 0.24$, satisfying the criterion of strong quantum fluctuation, $Q \sim 0.1$[58]. Strength of thermal fluctuation is evaluated by the Ginzburg number $G_i = (k_B T_c \gamma/\mu_0 H_c^2 \xi^3)^2/2$ [48,57], here $\mu_0 H_c = \mu_0 H_{c2}/\sqrt{2}\kappa$ is thermodynamic critical field, $\kappa = \lambda_{ab}/\xi_{ab}$ is the Ginzburg-Landau parameter, $\lambda_{ab}$ is the in-plane penetration depth. In high-$T_c$ superconductors with notable vortex liquid state, $G_i$ takes the order of $10^{-1}$ -$10^{-2}$, whilst conventional superconductors mostly have $G_i < 10^{-3}$ [56,59,60]. For Ba$_{0.75}$ClTaSe$_2$, we calculated $G_i$ using the parameters obtained from the $H_{c2}$ study and it yields $G_i \approx 2.2 \times 10^{-8} \kappa^4$. To fulfill the requirement of strong thermal fluctuation, this result hints at a value of $\kappa \sim 50$-100, comparable with typical copper-oxide and iron-based superconductors[60]. In this sense, Ba$_{0.75}$ClTaSe$_2$ is probably an extreme type-II superconductor characterized by large penetration depth and low superfluid density, the first phenomenological analogue of high-$T_c$ superconductors in the family of TMD-based materials. Further investigations are urgently needed to determine its superconducting pairing symmetry.

## CONCLUSIONS AND OUTLOOK

A comparison between $Ba_{0.75}ClTaS_2$ and $Ba_{0.75}ClTaSe_2$ provides more insights. Whereas the former has a stronger normal-state resistivity anisotropy (Figure S2), extremely enhanced superconducting anisotropy $\gamma$ (~150 versus ~20 in its counterpart) and $H_{c2}^{\|ab}$ are revealed for the latter (Table S3), wherein unique JV dynamics due to abnormally strong fluctuations is also observed. Given that the two compounds are isostructural, their electronic band structures are expected to be similar; hence, the distinct superconducting properties must be attributed to more subtle origins, such as the strength of spin-orbit interactions and electron correlations. Perhaps the larger SOC introduced by heavier Se atom plays a crucial role in provoking the novel behaviors in the superconducting state.

Our discoveries establish TMD-based superlattices as a promising new category of unconventional low-dimensional superconductors, which may create good opportunities for manipulating superconducting fluctuation and vortex dissipation. Last but not least, our method of synthesizing bulk TMD-based superlattices with $Ae$-Cl ($Ae$ is the alkaline earth metal) sublayers can be potentially applied to a variety of TMDs, adding to the strategies for regulating the physical properties of layered materials.

## ASSOCIATED CONTENT

**Supporting Information**.
The supporting information is available free of charge via the Internet at XXX.
Structure, transport and magnetization measurements; fitting models and parameters of the upper critical field; estimation of scattering time; resistivity anisotropy; transport properties in normal state; WAL analysis; Raman spectra; $I$-$V$ characteristics; determination of $T_{BKT}$; additional data of the upper critical field of $Ba_{0.75}ClTaS_2$; the upper critical field fitting for $Ba_{0.75}ClTaSe_2$; magnetization measurement of irreversibility field in $Ba_{0.75}ClTaSe_2$; definition of characteristic fields in $Ba_{0.75}ClTaSe_2$.; $H$-dependent in-plane resistivity measured in three different $Ba_{0.75}ClTaSe_2$ samples with $H$ applied parallel to the ab plane; refined crystal structures; fractional Atomic Coordinates and Equivalent Isotropic Displacement Parameters; superconducting transition temperature, upper critical fields and superconducting anisotropy in various TMD-based superconductors; the interlayer distances and coherence lengths; best-fit parameters in the two-band model fits.

**Accession Codes**
CCDC 2370523−2370524 contain the supplementary crystallographic data for this paper.

## AUTHOR INFORMATION

**Corresponding Author**

* Correspondence and requests for materials should be addressed to chenxh@ustc.edu.cn or zijixiang@ustc.edu.cn.

**Author Contributions**

‡These authors contributed equally: M.Z. Shi and K.B. Fan.

**Notes**

The authors declare no competing financial interest.

## ACKNOWLEDGMENTS

We are grateful to Jianjun Ying, Zhenyu Wang, Ye Yang and Min Shan for fruitful discussions. We thank Zhongliang Zhu, Yaolong Bian and Dongsheng Song for experimental assistance. This work was supported by the National Key Research and Development Program of the Ministry of Science and Technology of China (Grants Nos. 2022YFA1602601 and 2022YFA1602602), the National Natural Science Foundation of China (Grants Nos. 12488201, No. 12274390 and No. 12204448), the Innovation Program for Quantum Science and Technology (Grant No. 2021ZD0302802), the Anhui Initiative in Quantum Information Technologies (Grant No. AHY160000) and Chinese Academy of Sciences under Contract No. JZHKYPT-2021-08. J.Z. acknowledges the Natural Science Foundation of China (Grants No. 12122411). M.S. acknowledges the China Postdoctoral Science Foundation (2021M703107).

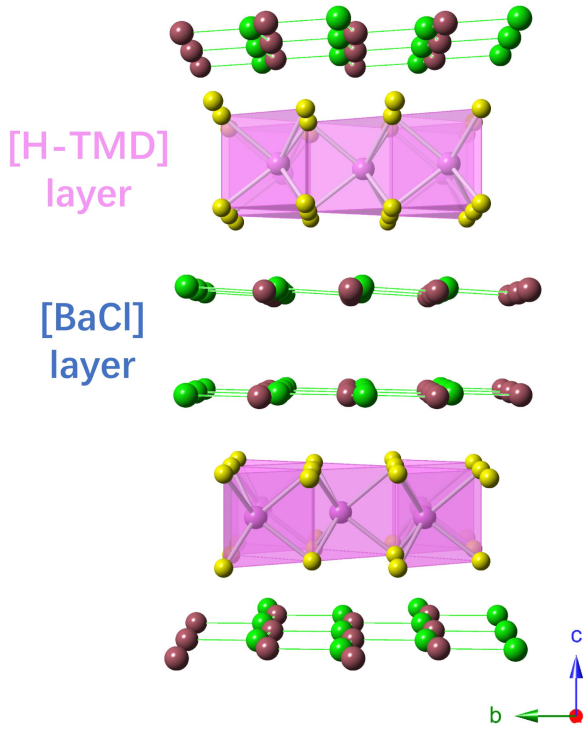
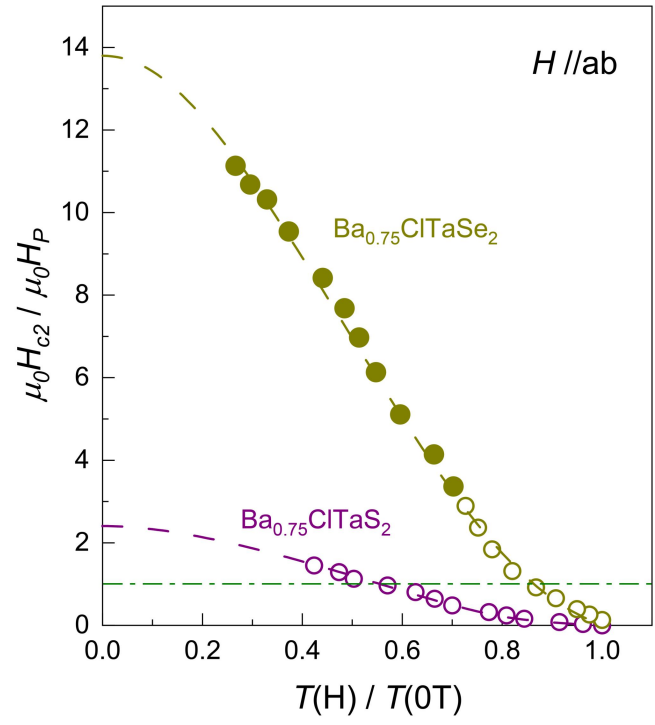